\documentclass{pasj00}
\begin{document}
\SetRunningHead{Imai et al.}{SiO Masers in the Galactic Nuclear Disk}
\Received{2001/11/14}%{yyyy/mm/dd}
\Accepted{2002/01/24}%{yyyy/mm/dd}

\title{Detections of SiO Masers from the Large-Amplitude Variables in the 
Galactic Nuclear Disk}

%%% begin:list of authors
\author{Hiroshi \textsc{Imai}$^{1,2}$, Shuji \textsc{Deguchi}$^{3}$, 
        Takahiro \textsc{Fujii}$^{2,4}$, Ian S. \textsc{Glass}$^{5}$, 
Yoshifusa \textsc{Ita}$^{6}$,}
\author{Hideyuki \textsc{Izumiura}$^{7}$, Osamu \textsc{Kameya}$^{1,2}$, 
        Atsushi \textsc{Miyazaki}$^{3}$, Yoshikazu \textsc{Nakada}$^{6}$
}

\and
\author{Jun-ichi {\sc Nakashima}$^{8}$}
%%%%%%%%%%%%%%%%%%% following is the note of the version %%%%
%\author{\\(will submitted to PASJ letter --- Version: 2001/07/15)\\}

\affil{$^{1}$ Mizusawa Astrogeodynamics Observatory, National 
Astronomical Observatory, Mizusawa, Iwate 023-0861}
\affil{$^{2}$ VERA Project Office, National Astronomical Observatory, 
              2-21-1 Osawa, Mitaka, Tokyo 181-8588}
\affil{$^{3}$ Nobeyama Radio Observatory, National Astronomical Observatory,
              Minamimaki, Minamisaku, Nagano 384-1305}
\affil{$^{4}$ Faculty of Science, Kagoshima University, 
              1-21-35 Korimoto, Kagoshima 890-0065}
\affil{$^{5}$ South African Astronomical Observatory,
              PO Box 9, Observatory 7935, South Africa}
\affil{$^{6}$ Institute of Astronomy, School of Science, The University of Tokyo,
              2-21-1 Osawa, Mitaka, Tokyo 181-0015}
\affil{$^{7}$ Okayama Astrophysical Observatory, National Astronomical 
Observatory, \\ Kamogata, Asakuchi, Okayama 719-0232}
\affil{$^{8}$ Department of Astronomical Science, The Graduate University 
for Advanced Studies,\\
Minamimaki, Minamisaku, Nagano 384-1305}	      
% \author{\\(will submitted to PASJ letter --- Version: 2001/07/15)\\}

%\affil{C-Address of Institute}\email{ccccc@xxx.xxx.xx.xx}
%%% end:list of authors

%%% Please use the following style in case that sorting by 
%%% affilation is impossible. 
%
% \author{%
%   D-Firstname \textsc{D-Familyname}\altaffilmark{1}
%   E-Firstname \textsc{E-Familyname}\altaffilmark{1,2}
%   and
%   F-Firstname \textsc{F-Familyname}\altaffilmark{2}}
% \altaffiltext{1}{Address of Institute}
% \email{ddddd@xxx.xxx.xx.xx}
% \email{eeeee@xxx.xxx.xx.xx}
% \altaffiltext{2}{Address of Institute}

%% `\KeyWords{}' always has to be placed before `\maketitle'.
 %Do NOT move this preamble from here!

\maketitle

\begin{abstract}
We have surveyed known large-amplitude variables within 15$'$ of the
Galactic center in the SiO $J=1$--0 $v=$ 1 and 2 maser lines at 43 GHz,
resulting in 79 detections and 58 non-detections. The detection rate of 58
percent is comparable to that obtained in Bulge IRAS source surveys.
SiO lines were also detected from four other sources near the program
objects. The SiO detection rate increases steeply with the period, particularly for
stars with $P>500$ d, where it exceeds 80\%. We found at a given period 
that the SiO detection rate is approximately double that for OH. 
These facts suggest that the large-amplitude variables in the Nuclear Disk region 
are AGB stars similar in their overall properties to the inner 
and outer bulge IRAS/SiO sources.
\end{abstract}

\section{Introduction}
The dynamical behavior of the central region of the Galaxy has attracted
much attention (\cite{mor96}). Since the bar-like structure of the
Galactic bulge was discovered (\cite{bli91}; \cite{nak91}; \cite{dwe95}), it
has been recognized that non-circular motions must be taken into account
when interpreting observational data, such as the CO gas distribution in
the central nuclear disk (\cite{bin91}).  Because double bars and nuclear
rings have been proposed as efficient mechanisms for feeding gas into the
centers of galaxies (\cite{shl89}), it has become additionally important to
look for signs of non-circularity in the motions of gas and stars. 

As a consequence of interstellar extinction, it is impossible to make
measurements of the radial velocities of stars in this region at visible
wavelengths. Most such information comes from radio observations of
SiO and OH masers which are not affected by interstellar extinction.  A
considerable amount of data is now available (\cite{lin92a}; \cite{deg00a}a,b).
The masers arise in the atmospheres of mass-losing stars on the Asymptotic
Giant Branch (AGB), which are intrinsically bright in the mid-infrared region, and
which can potentially be identified at these wavelengths.  However, the source
densities within one degree of the Galactic center are so high that the main
mid-infrared Catalog, the IRAS PSC, is very incomplete. Blind surveys in OH
and SiO maser lines for AGB stars have been made (\cite{sjo98};
\cite{shi97}; \cite{miy01}; \cite{deg02}), but they are limited in
sensitivity by short exposure times, and faint objects were difficult to detect 
compared with the pointed surveys. This situation, however, has recently changed 
owing to new ground-based surveys with near-infrared array cameras, 
such as 2MASS and DENIS, and to space-based mid-infrared surveys, such as MSX and ISOGAL 
(\cite{pri01}; \cite{omo99}). Large numbers of candidate stars suitable 
for maser surveys toward the nuclear disk have been discovered 
in the near-infrared $K$ band (\cite{gla01}) by making use of
their characteristic large-amplitude variability. 

In this paper, we report on preliminary results of an SiO maser survey of
Large Amplitude Variables (Miras or semiregulars; abbreviated as
LAV hereafter) within 15$'$ of the Galactic center (\cite{gla01}), whose
amplitudes and periods are known (\cite{woo98}; \cite{gla01}). Because these
stars are located at approximately the same distance (about 8 kpc) from the
Sun, they constitute an ideal sample for studying the statistical
characteristics of AGB stars and their detectability in the maser lines. In
addition, surveying these sources gives accurate radial velocities, and
provides basic data for investigating the kinematics of the Galactic nuclear
disk.

%\newpage

\section{Observations}
Simultaneous observations in the SiO $J=1$--0, $v=1$ and 2 transitions at
42.122 and 42.821 GHz, respectively, were made with the 45-m radio telescope
at Nobeyama during 2001 February--May. Details of SiO maser observations
using the NRO 45-m telescope have been described elsewhere (\cite{deg00a}a),
and are not repeated here.

The sources were taken from the Glass et al. (2001) list and were selected for
long periods ($P>450$ d) and good-quality light curves ($Q=$3). 
Additional LAV stars which do not satisfy the above criteria were also
chosen because of OH\,1612 MHz detections and positional coincidences
with MSX sources (\cite{pri01}). The positions of the stars were taken from
table 2 of \citet{gla01}, which are sufficiently accurate ($\sim$ a few
arsec) for a telescope beam size (HPBW) of about 40$''$.

In total, 134 LAVs were observed during the winter--spring season of the year
2001, and we detected SiO maser emission towards 76. According to the 
convention of \citet{gla01}, we write their names as the field and
star numbers, preceded by ``g'' for \citet{gla01}.   
We tried to observe all of the stars with
$Q=3$ and $P>450$ d in the \citet{gla01} sample, but two sources
(g23--3305 and g19--7) were left unobserved because of limited
observing time.
   
%%%%%%%%%%%%%%%%%%%%%%%%%%%%%%%%%%%%%% table 1 %%%%%%%%%%%%%%
\begin{table*}
  \caption{SiO radial velocities of the detected sources.}\label{tab:table1}
  \begin{tabular}{lr|lr|lr|lr}
  \hline\hline
Name & $V_{\rm ave}$ & Name & $V_{\rm ave}$ & Name & $V_{\rm ave}$ & Name & 
$V_{\rm ave}$\\
     & (km s$^{-1}$) &   & (km s$^{-1}$)&   & (km s$^{-1}$)&   & (km 
s$^{-1}$)\\
%\endfirsthead
	\hline
g1--8    &  79.0                  &g4--23   &   3.4  &g10--392 &  94.3$^{*}$           &g17--630 &  $-$9.8$^{*}$            \\
g1--31   & $-$38.6                &g4--113$^{\S}$&$-$279.0$^{3}$              &g11--4503 &  7.9$^{*}$      &g17--3762 & $-$49.0$^{*}$ \\
g1--72   &  21.1                  &g4--253  & $-$46.0       &g12--13  & $-$19.5            &g18--14  & $-$39.1        \\
g1--1890 &  23.6$^{*}$            &g4--557  & $-$85.1$^{*}$       &g12--21  & 33.3,71.5$^{5}$              &g19--2   & $-$17.8 \\
g2--1    &  $-$5.9$^{*}$          &g5--159  &  $-$2.4$^{*}$         &g12--51  &  44.4$^{*}$              &g19--7   & $-$107.1,$-$31.9$^{8}$ \\
g2--3    &  $-$7.5                &g5--2856 &  72.3$^{*}$         &g12--228 & $-$74.3          &g19--476  & $-$75.0$^{*}$   \\
g2--28   &  36.0$^{*}$            &g6--25$^{\S}$ & $-$0.8              &g13--16  &  83.2            &g19--685 & $-$53.5$^{9}$, $-$44.2$^{*}$ \\
g2--52   & 105.5                  &g6--85   &  88.1      &g13--18$^{\S}$  &  16.2              &g20--26$^{\S}$  & $-$38.4$^{*}$  \\
g2--697  & $-$52.5$^{*}$          &g6--135  &   5.8$^{*}$                &g13--33  & 124.8$^{*}$             &g20--136 & $-$95.4$^{10}$ \\
g2--6329 & 120.4$^{*}$            &g7--13   &   5.3               &g13--55  &  37.4          &g21--12  & $-$19.7                \\
g3--5    &35.9$^{1}$, 52.6$^{*}$  &g7--20   & $-$55.7              &g13--200&23.7$^{6}$         &g21--39$^{\S}$  & $-$65.1       \\
g3--6    &  22.9                  &g8--11   & $-$32.4      &g14--2$^{\S}$   & $-$58.6           &g22--4$^{\S}$   & $-$68.6$^{*}$  \\
g3--226  & 134.2$^{*}$            &g8--31$^{\S}$   & 108.5          &g14--6$^{\S}$   & $-$32.5  &g22--7   & $-$68.0          \\
g3--266  & $-$73.0$^{*}$          &g8--53   & $-$27.1         &g14--11  & $-$60.1                &g22--11  & $-$197.3 \\
g3--358  &  12.8$^{*}$            &g9--8    & $-$4.4, 72.3$^{4}$       &g16--1$^{\S}$   &  $-$5.2$^{7}$,22.1        &g22--22  & $-$11.4 \\
g3--779  &$-$142.3$^{*}$,$-$47.6$^{2}$  &g9--9$^{\S}$    &  23.1        &g16--8 & $-$38.9            &g22--76  & $-$77.3 $^{*}$ \\
g3--2752 &   5.7$^{*}$            &g10--5   &  70.0          &g16--25  & $-$61.2        &g22--274 & $-$85.6$^{11}$              \\
g3--2855 &$-$308.1$^{*}$          &g10--6   &  39.6$^{*}$     &g16--36$^{\S}$  &  50.2          &g23--5$^{\S}$   & $-$20.8       \\
g3--7655 &  $-$7.5                &g10--84  & $-$27.9$^{*}$           &g16--150$^{\S}$ &$-$131.5$^{*}$       &g23--8   & $-$109.0      \\
        
\hline
\end{tabular}
%\vspace{0.5cm}
\\
$^{*}$ The SiO radial velocity is consistent with that of OH 1612 MHz 
counterpart.\\
$^{\S}$  The star has a MSX counterpart (within 3 $\sigma$ of the MSX 
position uncertainty).\\
$^{1}$  the 35.9 km s$^{-1}$ component is from either OH 
359.958-0.058 (27$''$.2 away) with $V_{\rm lsr}=41.7$ km s$^{-1}$ or
OH 359.952$-$0.058 (32$''$.9 away) with $V_{\rm lsr}=36.4$ km s$^{-1}$. \\
$^{2}$  the $-$47.6 km s$^{-1}$ component is probably g3--1030$^{*}$ 
(OH 359.906$-$0.036,
20$''$.8 away).\\
$^{3}$  OH 359.855$-$0.078 at 3$''$.1 separation has a velocity at 4.8 
km s$^{-1}$.\\
$^{4}$  The 72.3 km s$^{-1}$ component is g9--547$^{*}$ (OH 0.040$-$0.056).\\
$^{5}$  Close to g12--65 and g12--42, definite assignment 
impossible.\\
$^{6}$  OH 0.178-0.055 at 5$''$.1 separation has $V_{\rm lsr}=-36.6$ km 
s$^{-1}$.\\
$^{7}$  the $-$5 km s$^{-1}$ component is OH 359.867+0.030 with 
$V_{\rm lsr}=-3.7$ km s$^{-1}$ (9$''$.6 
away).\\
$^{8}$  the g19--9 is located very close; definite assignment impossible.\\
$^{9}$  the $-$53.5 km s$^{-1}$ component is probably g19--128 which is 
located nearby.\\
$^{10}$  OH 359.791$-$0.081 at 0$''$.7  apart has a single peak at $-$116.8 
km s$^{-1}$, 
not inconsistent with the SiO velocity.\\
$^{11}$  OH 359.778+0.010 at 3$''$.3 apart has $V_{\rm lsr}=-27.5$ 
km s$^{-1}$. The $-85.6$ km s$^{-1}$ component may possibly be 
g22--1 at 20$''$.6 away, but this source was not detected 
in 2001 May (weather conditions poor).\\

\end{table*}

The detected sources and their radial velocities (average of the peak radial
velocities in the $J=1$--0 $v=1$ and 2 transitions) are given in table 1.  The
full line parameters (antenna temperatures, integrated intensities,
etc.), as well as null results, will be presented in a future paper.

The high stellar density near the Galactic center occasionally led to the
finding of an additional source within the 40$''$ telescope beam; we saw two
peaks in the spectra. The number of detections are, in fact, more than 76.
Table 1 lists 7 such sources.  
Most of these extra peaks can be assigned to nearby OH
sources or Glass et al's LAVs, except for two (g12--21 and g19-7). Because we
can easily find possible counterparts for each component in the doubly
peaked cases, we are confident that none of these spectra correspond to
single sources similar to the Orion IRc2 SiO masers (see the discussion in
\cite{deg99}). In the case of g9--8, we detected double peaks in the
spectra; the star g9--547 is close by (13$''$ separation). 
We took another SiO spectrum at a position 15$''$ offset towards g9--547, 
and confirmed that the relative intensities of the two peaks 
changed in the expected manner.
Thus, the identifications of these two sources are certain.

We checked the radial velocities of the 36 OH 1612 MHz maser counterparts
that are within 5$''$ of the positions of LAVs. The OH radial
velocities (the average of the OH double-peak velocities) of 32 counterparts
were consistent with the SiO radial velocities.  For these 32 sources, the
average SiO--OH velocity difference was $(V_{\rm SiO}-V_{\rm OH})_{\rm average}$=0.59
km s$^{-1}$ (with standard deviation of 1.93 km s$^{-1}$). In addition, we
found that one source, g20--136, has $V_{\rm lsr}=-85.6$ km s$^{-1}$, which is
not inconsistent with the single-peak OH detection at $V_{\rm lsr}=-116.8$ km
s$^{-1}$.

The SiO radial velocities of three stars were not
consistent with the OH velocities: g4--113 (OH 359.855$-$0.078), g13--200
(OH 0.178$-$0.055), and g22--274 (OH 359.791$-$0.081). The positions of two
sources coincide with the OH positions within 3$''$, the exception being
g13--200. We also looked at OH sources within 40$''$ of the SiO sources, but
found that none of them had similar radial velocities. These unidentified
SiO sources are possibly obscured AGB stars within the telescope beam or, 
in the case of OH 0.178$-$0.055, may be an AGB star that is not the OH
source; g13--200 is separated from OH 0.178$-$0.055 by 5$''$.1 (see table 6
of \cite{gla01}), which is too large to be a measurement error.

Twenty-three stars in the present sample have MSX source counterparts within
$3\sigma$ of the MSX positions (a few arcsec; \cite{pri01}).  They are
indicated by ``$\S$'' signs in table 1. The MSX subset contains 14
detections of SiO and 9 non-detections, giving a 61\% SiO detection rate.
The rate for the MSX counterparts is similar to the average detection rate, 
59\%, of SiO in the full sample.  These sources have flux densities of
0.3--4 Jy in the MSX band C ($\sim$12 $\mu$m) with colors not very different
from the IRAS/SiO sources in the Bulge (e.g., \cite{izu95}).

%%%%%%%%%%%%%%%%%%%%%%%%%%%%%%% section 3 %%%%%%%%%%%%%%%%

\section{Discussion}

\subsection{Period--SiO Detection Rate}
Figure 1 shows the period distribution of the observed sources and the SiO
detection rate (line graph). The sources with $P>450$ d (and $Q=3$) were
preferentially observed in the present survey so that the peak occurs in the
500--600 d bin. Note that the average period of the \citet{gla01} sample is
about 430 d. This figure clearly shows that the SiO detection rate increases
with the period. The dotted line in figure 1 indicates the OH detection rate,
which seems to correlate with the SiO detection rate quite well. 
The SiO detection rate, however, is twice as high as the OH detection rate, 
indicating that an SiO maser survey can double the number of stars 
with radial velocity data near the Galactic center.

That the maser detection rate increases with the period of light variation
in LAVs has already been suggested by observations of stars near the Sun
(e.g., \cite{ben96}). However, because of uncertainties in the distances of
these stars, the relation was not demonstrated clearly. In the present
sample, the distances to the LAV stars are almost equal and the correlation
with period is much more striking.

%%%%%%%%%%%%%%%%%%%%%%% figure 1 %%%%%%%%%%%%%%%%%%%%%%%%

\begin{figure}
  \begin{center}
    \FigureFile(80mm,50mm){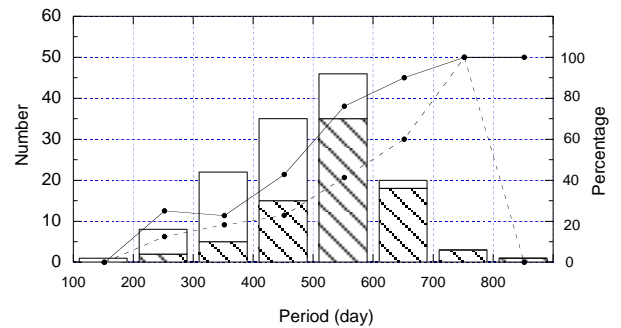}
    %%% \FigureFile(width,height){filename}
  \end{center}
  \caption{Histogram of period and detection probability (line graph). The 
  shadow and blank in the histogram indicate the SiO detection and 
nondetection, 
  respectively. The solid and broken lines (unit at the right vertical 
axis) 
  indicate the detection rate of SiO and
  OH masers, respectively, for the observed sources.}\label{fig:histogram}
\end{figure}
%%%%%%%%%%%%%%%%%%%%%%%%%%%%%%%%%%%%%%%%%%%%%%%%%

\subsection{Velocity Distribution}

Figure 2 shows longitude--velocity diagram for the detected sources.
Most of the radial velocities fall within $\pm$200 km s$^{-1}$. Two extreme
sources at $V_{\rm lsr}=-279$ and $-308$ km s$^{-1}$ occupy outlying positions.
These sources might be outer bulge objects in highly eccentric orbits, seen
by chance in the line of sight toward the Galactic Center [see discussion in
\citet{van92}].

The best fit for the radial velocities gives $V_{\rm lsr}=-19.2(\pm
$7.9)+0.085($\pm 0.020) (\Delta l/arcsec)$ km s$^{-1}$ and the standard
deviation from this line is 69.8 km s$^{-1}$ for the 79
sources. Here, $\Delta l$ is the Galactic longitude offset from 
Sgr A*, the dynamical center of the Galaxy 
(R.A.=$18^{\rm h}45^{\rm m}40^{\rm s}.05$, Dec.=$-29^{\circ}00'27''.9$,
J2000; \cite{rog94}). If we remove the two sources with
$|V_{\rm lsr}|>200$ km s$^{-1}$, the best fit gives $V_{\rm lsr}=-12.1(\pm
6.4)+0.078(\pm0.016) (\Delta l/arcsec)$ km s$^{-1}$. The best-fit slope,
0.085 km s$^{-1}$ per arcsec, which corresponds to 306 km s$^{-1}$ per degree,
is compatible with the value, $\sim$190 km s$^{-1}$ per degree, 
which was computed from the OH 1612 MHz data within 0.5 degree 
from the Galactic center (\cite{sjo98}). 
Considering the
slow rotation of the inner Bulge out to $3^{\circ}$ from the Galactic center,
$\sim$20 km s$^{-1}$ per degree (\cite{deg00a}a), we conclude that the
nuclear disk is rotating more rapidly than the inner Bulge.

The zero intercept at $\Delta l=0$ is negative, which is consistent with
previous SiO maser observations (\cite{izu95}; \cite{deg00a}a). This can be
interpreted as the motion of the local standard of rest toward the 
dynamical center of the Galaxy (\cite{izu95}).

The overall structure of the SiO $l$--$v$ diagram is quite similar to the OH
$l$--$v$ diagram in the same region (see figure 3 of \cite{sjo98}). We can
recognize a hole in the SiO distribution at ($\Delta l=$150$''$, $V_{\rm
lsr}=$ 80 km s$^{-1}$) in figure 2. The same hole can be seen in
figure 3 of \citet{sjo98}, though another hole at ($\Delta l=$-200$''$, $V_{\rm
lsr}=$ 0 km s$^{-1}$) in figure 3 of \citet{sjo98} is not present in figure
2 of the present paper. It is known that a bar potential produces a
characteristic pattern in the $l$--$v$ diagram (\cite{kui95}). Non-circular
streaming motions of stars may be present in the nuclear disk region in our
Galaxy, and the hole in the $l$--$v$ diagram can be a signature of
bars within the bar, such as have been found in other galaxies
(\cite{erw01}). An alternative interpretation is that the stars
within 100$''$ from the Galactic center have a large circular motion
($\sim 1.1$ km s$^{-1}$ per arcsec; \cite{deg02}). The magnitude
of the circular motion is comparable with that of the
circumnuclear ring revealed by the molecular line observations
(e.g., \cite{wri01}). It is possible that the rapidly rotating 
AGB stars near the Galactic center were born as a result of
star formation in the circumnuclear ring (\cite{lev95}). 

%%%%%%%%%%%%%%%%%%%%%%% figure 2 %%%%%%%%%%%%%%%%%%%%%%%%
\begin{figure}
  \begin{center}
    \FigureFile(80mm,50mm){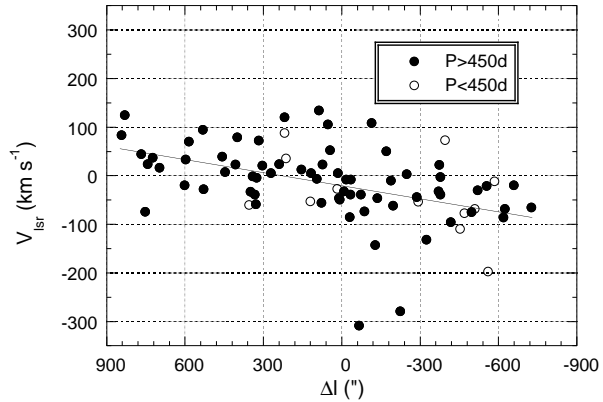}
    %%% \FigureFile(width,height){filename}
  \end{center}
  \caption{Longitude--Velocity diagram for the detected sources. 
The filled and open circles indicate the stars with periods above 
and below 450 d, respectively.
  The best linear fit is shown by the solid line.}\label{fig:l-v.diagram}
\end{figure}
%%%%%%%%%%%%%%%%%%%%%%%%%%%%%%%%%%%%%%%%%%%%%%%%%%%%%%%%%%%%%%%%%%%%%%%%%%%
%%%

\section{Conclusion}

We surveyed 134 large-amplitude variables within 15$'$ of the Galactic
center, and obtained 79 detections in the SiO maser lines. The SiO
detection rate of $\sim$60\% is comparable with the previous SiO survey of
color-selected bulge IRAS sources. The SiO detection rate increases with
the period of the light variation, and is well correlated with the OH
detection rate.  The longitude-velocity diagram of the SiO sources has been
revealed to be quite similar to the OH $l$--$v$ diagram. These facts suggest
that these large-amplitude variables in the Galactic nuclear disk are
mass-losing stars in the AGB phase, quite similar to the IRAS sources in the
inner Galactic Bulge.

This research was partly supported by Scientific Research Grant
(C2) 12640243 of Japan Society for Promotion of Sciences. 
%Authors thank Prof. H. Habing for useful comments.

%%%%%%%%%%%%%%%%%%%%%%%%%%%%%%%%%%%%%%%

%%%

%%%%%%%%%%%%%%%%%%%%%%%%%%%%%%%%%%%%%%%%%%%%%%%%%%%%%%%%%%%
\end{document}